\documentclass[11pt]{article}
\usepackage{ifthen} 
\newboolean{pdflatex}
\setboolean{pdflatex}{true} 
\newboolean{articletitles}
\setboolean{articletitles}{true} 

\newboolean{uprightparticles}
\setboolean{uprightparticles}{false} 
\usepackage[a4paper, portrait, margin=2.5cm]{geometry}
\usepackage{graphicx}
\usepackage{fancyhdr}
\usepackage{layout}
\usepackage{hyperref} 

\newcommand{\lhcborcid}[1]{\href{https://orcid.org/#1}{\hspace*{0.1em}\raisebox{-0.45ex}{\includegraphics[width=1em]{figs/orcidIcon.pdf}}}}

\newcommand{\AS}{{\ensuremath{A_{S}}}\xspace}

\newcommand{\gevt}{\ensuremath{\mathrm{Ge\kern -0.1em V}}\xspace}

\newcommand{\AAv}[1][\Ai]{#1^{\text{CP}}}

\newcommand{\magSAv}[1][]{|\AAv[\AS]|}
\newcommand{\magSAvSq}[1][]{\magSAv^2}

\usepackage{xspace} 
\usepackage{upgreek}

\def\MagUp {\mbox{\em Mag\kern -0.05em Up}\xspace}

\ifthenelse{\boolean{uprightparticles}}%
{

 \def\PDelta      {\ensuremath{\Delta}\xspace}                 
 \def\PXi         {\ensuremath{\Xi}\xspace}                 
 \def\PLambda     {\ensuremath{\Lambda}\xspace}                 
 \def\PSigma      {\ensuremath{\Sigma}\xspace}                 
 \def\POmega      {\ensuremath{\Omega}\xspace}                 
 \def\PUpsilon    {\ensuremath{\Upsilon}\xspace}
 \let\oldPi\Pi
 \def\PPi         {\ensuremath{\oldPi}\xspace}

 \def\PB      {\ensuremath{\mathrm{B}}\xspace}                 
                  
 \def\PD      {\ensuremath{\mathrm{D}}\xspace}

 \def\PK      {\ensuremath{\mathrm{K}}\xspace}

 \def\Pi      {\ensuremath{\mathrm{i}}\xspace}

 \def\Ps      {\ensuremath{\mathrm{s}}\xspace}

 \def\thebaroffset{0.0em}
}
{

 \mathchardef\PDelta="7101
 \mathchardef\PXi="7104
 \mathchardef\PLambda="7103
 \mathchardef\PSigma="7106
 \mathchardef\POmega="710A
 \mathchardef\PUpsilon="7107
 \mathchardef\PPi="7105
                  
 \def\PB      {\ensuremath{B}\xspace}                 
                  
 \def\PD      {\ensuremath{D}\xspace}

 \def\PK      {\ensuremath{K}\xspace}

 \def\Pi      {\ensuremath{i}\xspace}

 \def\Ps      {\ensuremath{s}\xspace}

 \def\thebaroffset{0.18em}
}
\newcommand{\offsetoverline}[2][\thebaroffset]{\kern #1\overline{\kern -#1 #2}}%

\makeatletter
\ifcase \@ptsize \relax
  \newcommand{\miniscule}{\@setfontsize\miniscule{4}{5}}
\or
  \newcommand{\miniscule}{\@setfontsize\miniscule{5}{6}}
\or
  \newcommand{\miniscule}{\@setfontsize\miniscule{5}{6}}
\fi
\makeatother

\DeclareRobustCommand{\optbar}[1]{\shortstack{{\miniscule (\rule[.5ex]{1.25em}{.18mm})}
  \\ [-.7ex] $#1$}}

\def\squark    {{\ensuremath{\Ps}}\xspace}

\def\KorKbar {\kern \thebaroffset\optbar{\kern -\thebaroffset \PK}{}\xspace}

\def\D       {{\ensuremath{\PD}}\xspace}

\def\DorDbar {\kern \thebaroffset\optbar{\kern -\thebaroffset \PD}\xspace}

\def\Dp      {{\ensuremath{\D^+}}\xspace}
\def\Dm      {{\ensuremath{\D^-}}\xspace}

\def\DpDm    {\ensuremath{\Dp {\kern -0.16em \Dm}}\xspace}

\def\B       {{\ensuremath{\PB}}\xspace}

\def\BorBbar {\kern \thebaroffset\optbar{\kern -\thebaroffset \PB}\xspace}

\def\Bd      {{\ensuremath{\B^0}}\xspace}

\def\BdorBdbar {\kern \thebaroffset\optbar{\kern -\thebaroffset \Bd}\xspace}

\def\Bs      {{\ensuremath{\B^0_\squark}}\xspace}

\def\BsorBsbar {\kern \thebaroffset\optbar{\kern -\thebaroffset \Bs}\xspace}

\def\Y#1S{\ensuremath{\PUpsilon{(#1S)}}\xspace}

\def\LorLbar     {\kern \thebaroffset\optbar{\kern -\thebaroffset \PLambda}\xspace}

\def\to                 {\ensuremath{\rightarrow}\xspace}

\def\CP                {{\ensuremath{C\!P}}\xspace}

\def\AT#1     {\ensuremath{A_{\mathrm{T}}^{#1}}\xspace}           

\def\C#1      {\ensuremath{\mathcal{C}_{#1}}\xspace}                       
\def\Cp#1     {\ensuremath{\mathcal{C}_{#1}^{'}}\xspace}                    
\def\Ceff#1   {\ensuremath{\mathcal{C}_{#1}^{\mathrm{(eff)}}}\xspace}        
\def\Cpeff#1  {\ensuremath{\mathcal{C}_{#1}^{'\mathrm{(eff)}}}\xspace}       
\def\Ope#1    {\ensuremath{\mathcal{O}_{#1}}\xspace}                       
\def\Opep#1   {\ensuremath{\mathcal{O}_{#1}^{'}}\xspace}                    

\newcommand{\aunit}[1]{\ensuremath{\text{\,#1}}}       

\newcommand{\tev}{\aunit{Te\kern -0.1em V}\xspace}
\newcommand{\gev}{\aunit{Ge\kern -0.1em V}\xspace}
\newcommand{\mev}{\aunit{Me\kern -0.1em V}\xspace}
\newcommand{\kev}{\aunit{ke\kern -0.1em V}\xspace}
\newcommand{\ev}{\aunit{e\kern -0.1em V}\xspace}
 
\newcommand{\mevc}{\ensuremath{\aunit{Me\kern -0.1em V\!/}c}\xspace}
\newcommand{\gevc}{\ensuremath{\aunit{Ge\kern -0.1em V\!/}c}\xspace}
\newcommand{\mevcc}{\ensuremath{\aunit{Me\kern -0.1em V\!/}c^2}\xspace}
\newcommand{\gevcc}{\ensuremath{\aunit{Ge\kern -0.1em V\!/}c^2}\xspace}

\def\gsim{{~\raise.15em\hbox{$>$}\kern-.85em
          \lower.35em\hbox{$\sim$}~}\xspace}
\def\lsim{{~\raise.15em\hbox{$<$}\kern-.85em
          \lower.35em\hbox{$\sim$}~}\xspace}

\def\tell1  {TELL1\xspace}
\def\ukl1   {UKL1\xspace}

\usepackage{microtype}
\usepackage{lineno}  
\usepackage{xspace} 
\usepackage{caption} 

\usepackage{graphicx}  
\usepackage{color}
\usepackage{colortbl}
\graphicspath{{./figs/}} 

\usepackage{amsmath} 
\usepackage{amssymb}
\usepackage{amsthm}
\usepackage{amsfonts}
\usepackage{upgreek} 

\newcommand*\patchAmsMathEnvironmentForLineno[1]{%
\expandafter\let\csname old#1\expandafter\endcsname\csname #1\endcsname
\expandafter\let\csname oldend#1\expandafter\endcsname\csname
end#1\endcsname
 \renewenvironment{#1}%
   {\linenomath\csname old#1\endcsname}%
   {\csname oldend#1\endcsname\endlinenomath}%
}
\newcommand*\patchBothAmsMathEnvironmentsForLineno[1]{%
  \patchAmsMathEnvironmentForLineno{#1}%
  \patchAmsMathEnvironmentForLineno{#1*}%
}
\AtBeginDocument{%
\patchBothAmsMathEnvironmentsForLineno{equation}%
\patchBothAmsMathEnvironmentsForLineno{align}%
\patchBothAmsMathEnvironmentsForLineno{flalign}%
\patchBothAmsMathEnvironmentsForLineno{alignat}%
\patchBothAmsMathEnvironmentsForLineno{gather}%
\patchBothAmsMathEnvironmentsForLineno{multline}%
\patchBothAmsMathEnvironmentsForLineno{eqnarray}%
}

\usepackage{hyperxmp}

\usepackage[colorinlistoftodos,textsize=scriptsize]{todonotes}

\usepackage[bottom,flushmargin,hang,multiple]{footmisc}

\usepackage[all]{hypcap} 

\usepackage{cite} 
\usepackage{mciteplus}
\usepackage{footmisc}

\usepackage{mciteplus}
\usepackage{footmisc}
\usepackage{longtable} 
\usepackage{booktabs} 
\usepackage{rotating} 
\usepackage{lscape}
\usepackage{units}
\usepackage{overpic}
\usepackage{rotating} 
\usepackage{subfig}
\usepackage{pbox}
\usepackage{tabularx}
\usepackage{multirow}
\usepackage{multicol}
\usepackage{listings}
\usepackage{color}
\usepackage[none]{hyphenat}
\usepackage{amsmath}

\fancypagestyle{firstpage}{%
  \setlength\headheight{154pt}
  \fancyhf{}%
  \fancyhead[C]{\includegraphics[width=\textwidth]{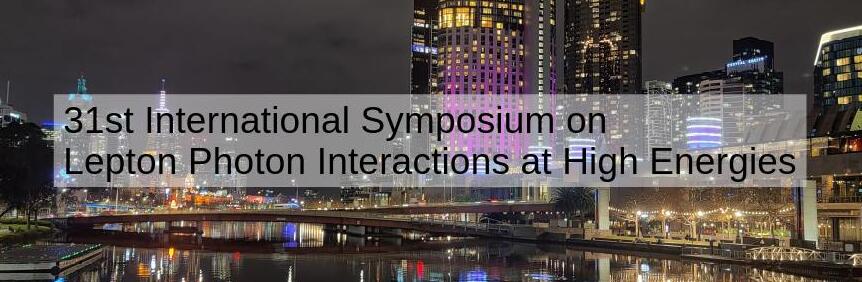}}%
  \fancyfoot[C]{CC-BY-4.0 licence}%
}
\begin{document}

\title{\CP violation in heavy-flavour decays} 
\date{17-21 July 2023}
\author{Yuehong Xie \\ Central China Normal University \\ 
\textit{On behalf of the LHCb collaboration}} 

\newgeometry{top=2cm, bottom=7cm}
\maketitle
\thispagestyle{firstpage}
\begin{abstract}
The   CKM matrix is the only source of \CP violation  in the Standard Model. Heavy-flavour decays provide an ideal laboratory to test the CKM mechanism. Some recent highlights in \CP violation in heavy-flavour decays from the LHCb and  Belle II  experiments are presented, including updated measurements of the CKM angles $\beta$, $\gamma$ and $\phi_s$, and new results in \CP violation in charm decays.
\end{abstract}
\newpage
\restoregeometry

\section{Introduction}

The fact that charge-parity (\CP) symmetry is broken in weak interactions has been known for more than sixty years. The discovery of large \CP violation in $B^0$ decays by BaBar~\cite{BaBar:2001pki} and Belle~\cite{Belle:2001zzw} in 2001
established the  CKM mechanism~\cite{Cabibbo:1963yz, Kobayashi:1973fv} as the leading source of \CP violation in the Standard Model (SM). In this theory, quark flavour transitions are fully described by the $3\times 3$  unitary CKM matrix using  four parameters, including one irreducible complex phase responsible for \CP violation, which are determined from flavour physics data. Study of heavy-flavour decays can access most  CKM matrix elements, providing crucial information for precision test of the CKM mechanism. 
The LHCb and Belle II experiments are playing dominant roles in this field.

The LHCb detector~\cite{LHCb:2008vvz} is a forward spectrometer at the Large Hadron Collider, dedicated to  the study of $b$- and $c$-hadron decays.  It has excellent performance in tracking, vertexing and particle identification, and provides an effective $b$-tagging efficiency of about 5\% and a typical time resolution of about 45 fs~\cite{LHCb:2014set}, sufficient for resolving the fast \Bs oscillation.  
LHCb has collected large samples of proton-proton  collisions, corresponding to an integrated luminosity of 3 ${\rm fb}^{-1}$ at a center-of mass energy of 7 and 8 TeV in 2011--2012 (Run 1) and  and 6  ${\rm fb}^{-1}$ at 13 TeV in 2015--2018 (Run 2). Based on these data, LHCb has performed extensive studies of  beauty and charm decays, and achieved high-precision measurements of \CP-violating parameters~\cite{Chen:2021ftn}. This paper summarizes the recent progress  using the full LHCb Run 1 and Run 2 data samples. 

The Belle II experiment ~\cite{Belle-II:2010dht} is a super-$B$ factory at SuperKeKB. The Belle II physics program is complementary to that of the LHCb experiment, with unique strengths to study $B$ decays to final states with neutrinos and neutral particles. 
It provides an effective $b$-tagging efficiency of about 30\% and a typical time resolution of about 1.5 ps for study of \Bd mixing and \CP violation. Since the start of operation in 2019, Belle II has collected $e^+e^-$ collision  data  at the $\Upsilon$(4S) energy corresponding to  an integrated luminosity of more than 360 ${\rm fb}^{-1}$. This paper discusses some  \CP violation measurements  based on this initial data sample.


\section{Measurement of $\sin2\beta$}

The CKM angle 
\mbox{$\beta  \equiv \rm{arg}\left(-V_{\textit{cd}}^{\phantom{\ast}}V_{\textit{cb}}^{\ast}/
V_{\textit{td}}^{\phantom{\ast}}V_{\textit{tb}}^{\ast}\right)$} is a key parameter to characterize mixing-induced \CP asymmetry in \Bd decays,
where $V_{qq^{\prime}}$ represents the relevant CKM matrix element.
The time-dependent  \CP asymmetry   in \Bd decays via a  $b\to c\bar{c}s$
transition to a \CP eigenstate with eigenvalue $\eta_f $ follows the relation $A_{\rm CP}(t) \approx \eta_f \sin2\beta \sin(\Delta m_d t)  $, where  the approximation assumes no \CP violation in the mixing and decay.  The  effective value of $2\beta$  is very sensitive to new physics in $B^0$-$\bar{B}^0$ mixing,
making it a  flagship measurement
for heavy-flavour experiments.
Both Belle and BaBar experiments have achieved a precision better than 0.03  combining  several  $b\to c\bar{c}s$ decay modes ~\cite{BaBar:2009by,Belle:2012paq}. LHCb measured $\sin2\beta=0.760\pm 0.034$ in $\Bd \to J/\psi K^0_S$  and $\Bd \to \psi(2S) K^0_S$  decays
using Run 1 data~\cite{LHCb-PAPER-2017-029}, compatible with  the world average value  $\sin2\beta=0.699\pm 0.017$~\cite{HFLAV21}. 

\begin{figure}[tbp]
\centering
\includegraphics[width=.44\textwidth]{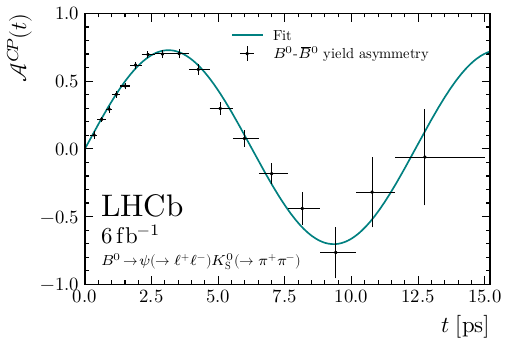}
\includegraphics[width=.4\textwidth]{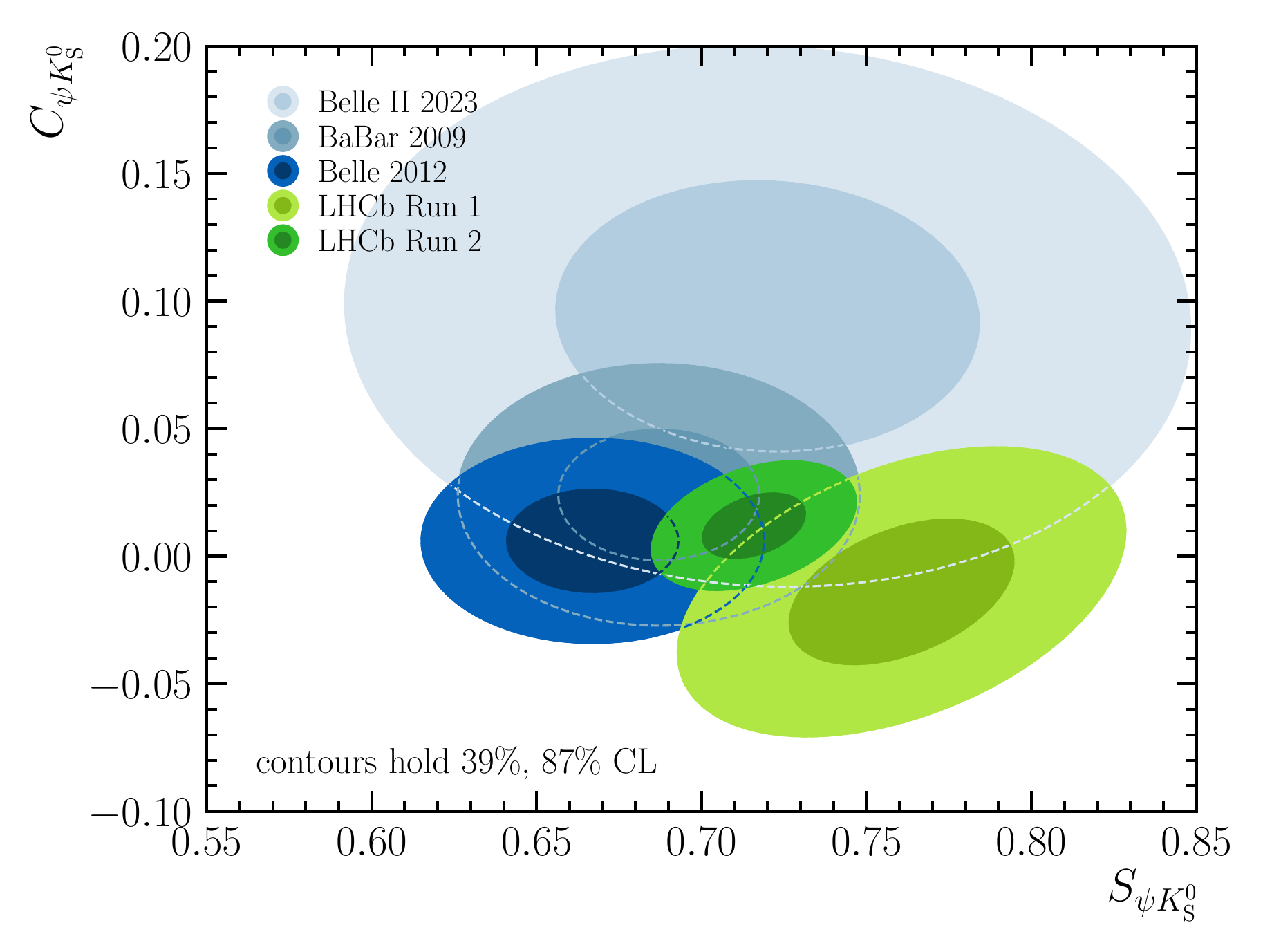}
\caption{\small Left: asymmetry between $B^{0}$- and $\bar{B}^0$-tagged decays 
 as a function of the decay time obtained using LHCb Run 2 data, overlaid with the fit projection; right: comparison of the two-dimensional contours 
in the $C$ versus $S$ plane from  LHCb, BaBar, Belle and Belle II. 
}
\label{fig:asymmetry1}
\end{figure}

Recently, LHCb updated the $\sin2\beta$ measurement 
using the Run 2 data sample. The  measured time-dependent \CP asymmetry and the fit projection are shown on the left of Figure~\ref{fig:asymmetry1}. 
The direct \CP violation parameter, $C$,  is found to be consistent with zero,
 and mixing-induce \CP violation is measured to be  $S =\sin2\beta=0.717\pm 0.013 (\rm stat) \pm 0.008 (\rm syst)$~\cite{LHCb-PAPER-2023-013}. 
   A combination of the LHCb Run 1 and Run 2 results yields $\sin2\beta = 0.724 \pm 0.014$.
With Run 2 data,  LHCb  overtakes the $B$ factories in the precision of  $\sin2\beta$ measurement (Figure~\ref{fig:asymmetry1} right),  and improves the precision of the world average by 35\%. The new world average, 
 $\sin2\beta = 0.708 \pm 0.011$,
is  consistent with the indirect determinations  by  the CKMfitter group~\cite{CKMfitter2015} and UTfit group~\cite{UTfit-UT}. LHCb plans to accumulate 300 ${\rm fb}^{-1}$ of proton-proton collisions  
after a phase-II upgrade, and improve the precision of $\sin2\beta$  to 0.003~\cite{LHCb:2018roe}.

Using the  $\Upsilon$(4S)  data  collected in 2019--2021 and corresponding to 190 ${\rm fb}^{-1}$, Belle II   has measured the \Bd  mass difference 
$\Delta m_d = 0.516 \pm 0.008 (\rm stat) \pm 0.004 (\rm syst)\; {\rm ps}^{-1}$ in $\Bd \to D^{(*)-}\pi^+$ decays ~\cite{Belle-II:2023bps}, and the \CP violation parameter  
$\sin2\beta=0.720 \pm 0.062 (\rm stat) \pm 0.016 (\rm syst)$ in the decay $\Bd \to J/\psi K^0_S$ ~\cite{Belle-II:2023nmj}. 
Belle II has also reported measurements of effective $\sin2\beta$ value in penguin-dominated $b\to s$ transitions, such as $\Bd \to \phi K^0_S$~\cite{Belle-II:2023oud}, $\Bd \to K^0_S \pi^0$~\cite{Belle-II:2023grc} and $\Bd \to K^0_S K^0_S K^0_S$~\cite{Belle-II:2022nof}.
While being statistically limited, these early measurements  demonstrate Belle II's ability to study \Bd mixing and \CP violation. It plans to collect  50 ${\rm ab}^{-1}$  at the $\Upsilon$(4S) energy, which will enable the key parameter $\sin2\beta$   to be measured at a precision of 0.005~\cite{Belle-II:2010dht}.

\section{Measurement of \Bs mixing phase $\phi_s$}

The \Bs mixing phase $\phi_s$ is the counterpart of the \Bd mixing phase $2\beta$. Its value is precisely predicted to be very small in the SM, $\phi_s^{\rm SM} =-2\beta_s=-0.0368 ^{+0.0009}_{-0.0006}$ rad~\cite{CKMfitter2015}, 
where \mbox{$\beta_s  \equiv 
 - \rm{arg}\left(-V_{\textit{cb}}^{\phantom{\ast}}V_{\textit{cs}}^{\ast}/
V_{\textit{tb}}^{\phantom{\ast}}V_{\textit{ts}}^{\ast}\right)$}, and sensitive to new physics in \Bs-$\bar{B}^0_s$ mixing.
 The $\phi_s$ angle can be accessed in time-dependent \CP asymmetries in \Bs decays to 
\CP eigenstates via $b\to c\bar{c}s$ transitions. 
The decay $\Bs \to J/\psi \phi$ is the golden mode to measure $\phi_s$. A flavour-tagged time-dependent analysis is required to determine the mixing-induced \CP asymmetry $S=\eta_f\sin\phi_s$.
In addition, an angular analysis is needed to statistically separate the  \CP-even and odd components of the $ J/\psi \phi$ system. The
LHCb, ATLAS and CMS experiments 
have all measured the $\phi_s$ parameter and the \Bs decay width parameter, $\Delta\Gamma_s$, in  $\Bs \to J/\psi \phi$ decays using  Run 1 and part of Run 2 data~\cite{LHCb-PAPER-2019-013, ATLAS:2020lbz,CMS:2020efq}. While the $\phi_s$ values agree well, there is some tension between the three measurements of $\Delta\Gamma_s$.
Combining these results with  LHCb measurements in other $b\to c\bar{c}s$ modes leads to the world average $\phi_s=-0.049 \pm 0.019$ rad~\cite{HFLAV21}.

\begin{figure}[tbp]
\centering
\includegraphics[width=.44\textwidth]{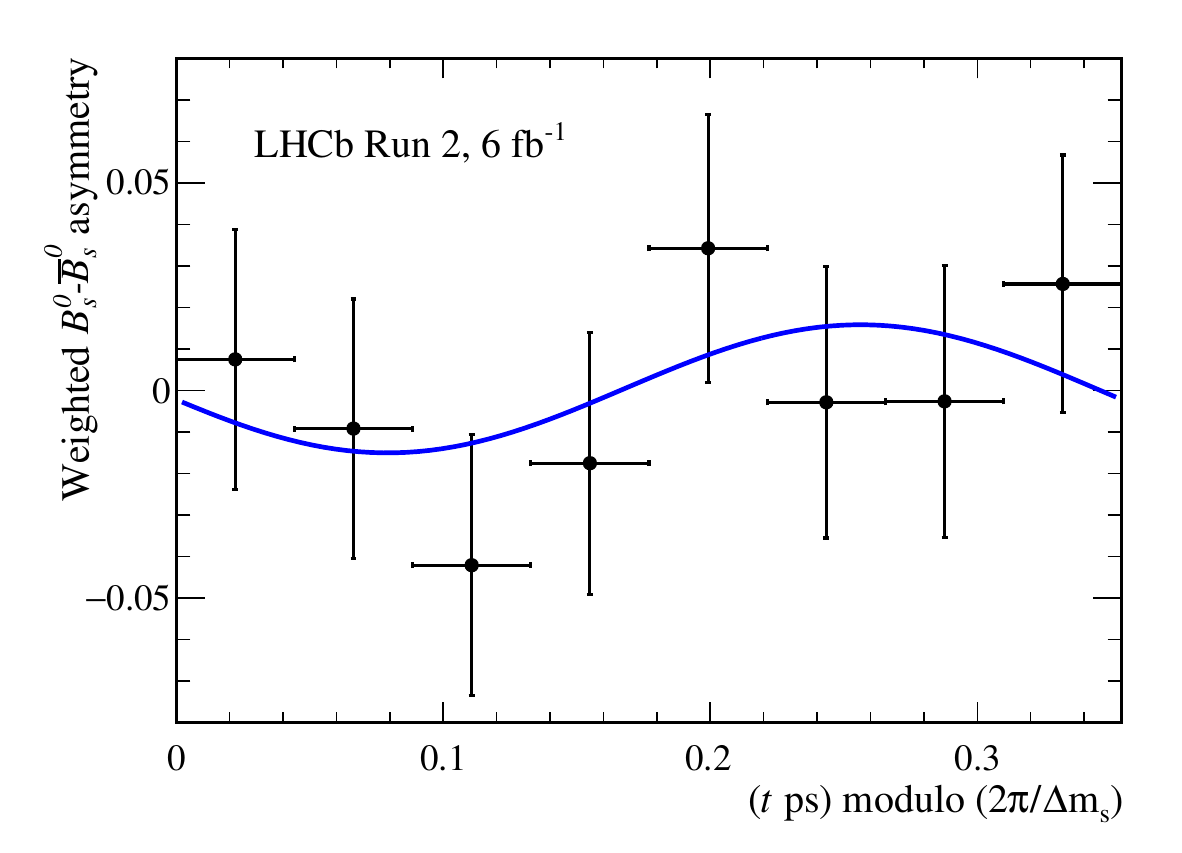}
\includegraphics[width=.45\textwidth]{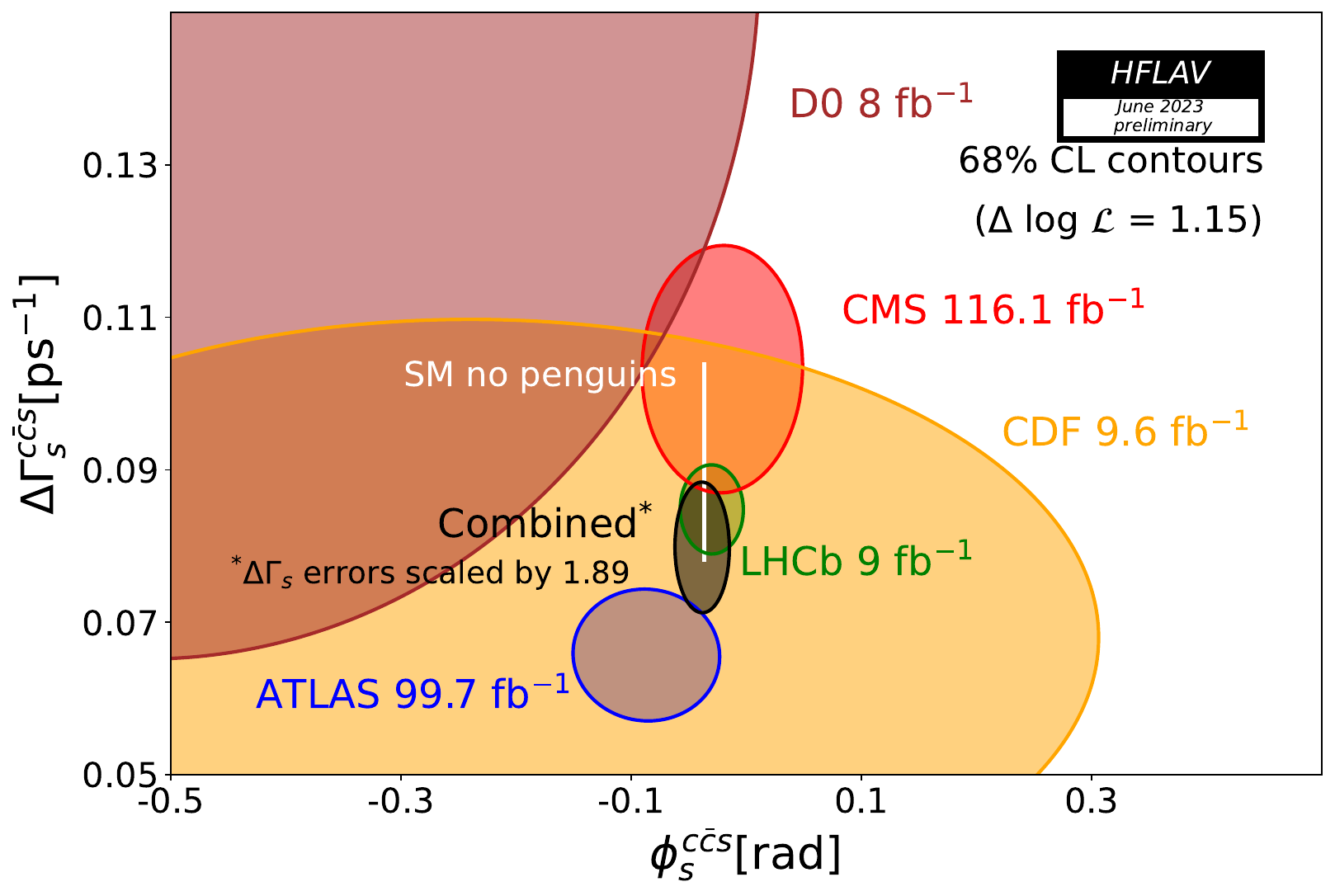}
\caption{\small Left: asymmetry between $B_s^{0}$- and $\bar{B}_s^0$-tagged decays as a function of the decay time obtained using LHCb Run 2 data, overlaid with the fit projection; right: comparison and combination of measurements from different experiments.
}
\label{fig:asymmetry}
\end{figure}

Recently, LHCb reported an improved measurement of $\phi_s$
obtained using the full Run 2 data sample in the decay $\Bs \to J/\psi \phi$~\cite{LHCb-PAPER-2023-016}.
The left plot in Figure~\ref{fig:asymmetry} shows the measured time-dependent \CP asymmetry.
No evidence for \CP violation is observed, and the parameters are measured to be $\phi_s = -0.039 \pm 0.022(\rm stat) \pm 0.006 (\rm syst)$ rad
and $\Delta \Gamma_s =0.0845 \pm 0.0044 (\rm stat) \pm 0.0024 (\rm syst)\; {\rm ps}^{-1}$. A combination with previous independent LHCb measurements in $b\to c\bar{c}s$ decays
yields the LHCb average value, $\phi_s =-0.031 \pm 0.018$ rad. This new result improves the precision of  the world average by 15\%. As seen in the right plot of Figure~\ref{fig:asymmetry},
the new world average, $\phi_s =-0.039 \pm 0.016$ rad, is dominated by the  LHCb measurement and consistent with the indirect determinations~\cite{CKMfitter2015, UTfit-UT}. Looking to the future, LHCb aims to achieve  a precision of 0.003 rad for $\phi_s$   
after the phase II upgrade
 using a data sample of 300 ${\rm fb}^{-1}$~\cite{LHCb:2018roe}.

LHCb has also studied time-dependent \CP violation in the penguin-dominated decay   $\Bs \to \phi \phi$, which proceeds via a $b\to s\bar{s}s$ transition and can probe quantum effects of new particles in the loop decay process. Previously, LHCb  measured the \CP-violating phase, $\phi_s^{s\bar{s}s} = -0.073 \pm 0.115 (\rm stat) \pm 0.027 (\rm syst)$ rad~\cite{LHCb-PAPER-2019-019},  using  data taken in 2011--2017. A recent update of this analysis using the full Run 1 and Run 2 data samples leads to the improved measurement,  $\phi_s^{s\bar{s}s} = -0.074 \pm 0.069$ rad~\cite{LHCb-PAPER-2023-001}. This is the most precise measurement of mixing-induced \CP asymmetry in any penguin-dominated $B$ decay to date, 
and agrees with the SM prediction of tiny \CP violation~\cite{Raidal:2002ph,Cheng:2009mu,Wang:2017rmh}.
The \CP-violating parameters are also measured separately for  different polarization states of the $\phi\phi$ pair, and no sign of polarization dependence is observed.

\section{Measurement of CKM angle $\gamma$}

The CKM angle \mbox{$\gamma \equiv \textrm{arg}(-V_{ud}V^*_{ub}/V_{cd}V_{cb}^*)$} can be measured through interference of $b\to u$ and $b \to c$ quark transitions in purely tree-level decays. These measurements are free of new physics effect and provide a SM benchmark, which can be compared to indirect determinations 
inferred from measurements of other observables sensitive to new physics.
Any significant discrepancy will be a clear sign of physics beyond the SM. The golden modes to measure $\gamma$ are $B \to D h$ decays, where $h$ is a charged kaon or pion meson,  and $D$ is a superposition of $D^0$ and $\bar{D}^0$ mesons, which can be reconstructed in many different modes. Other $B$ decay modes, such as $B\to D^{*0} h$, are also used to measure $\gamma$. 

\begin{figure}[htb]
\begin{center}
\includegraphics[width=0.45\linewidth]{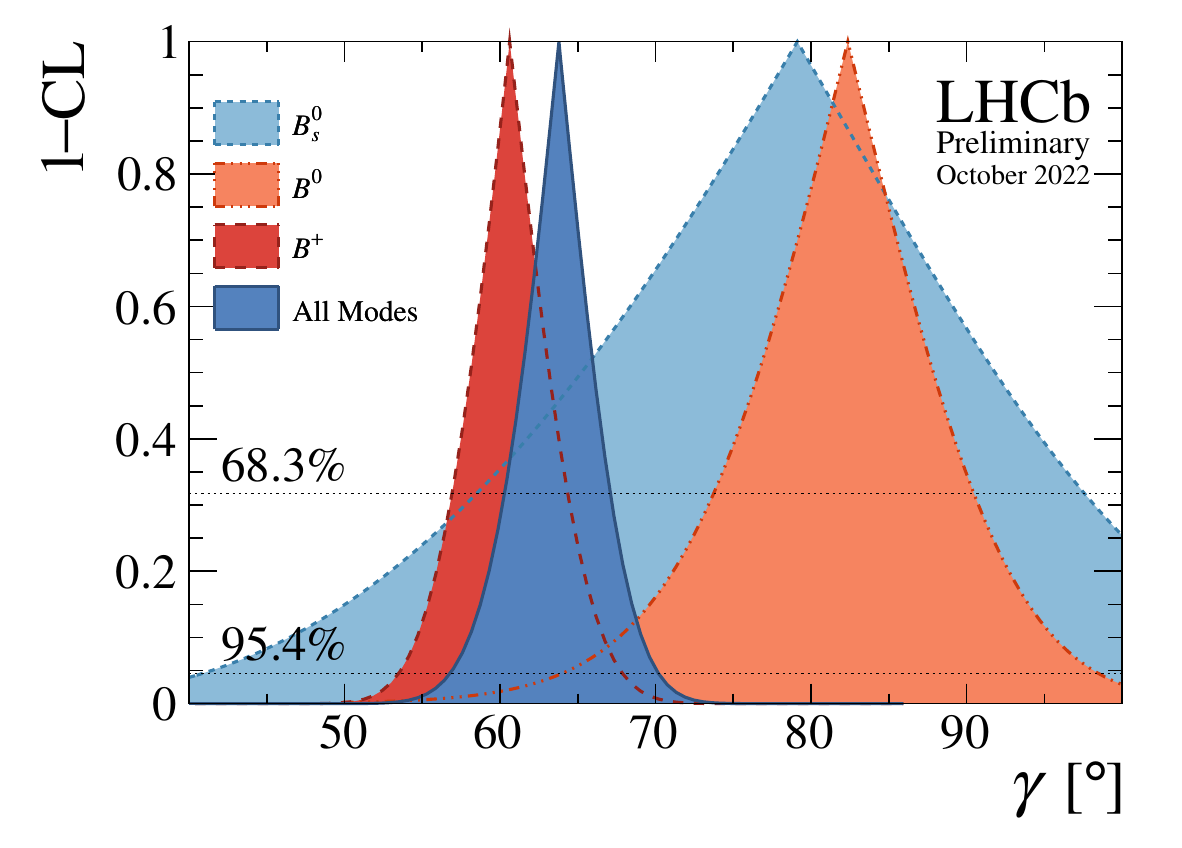}
\includegraphics[width=0.46\linewidth]{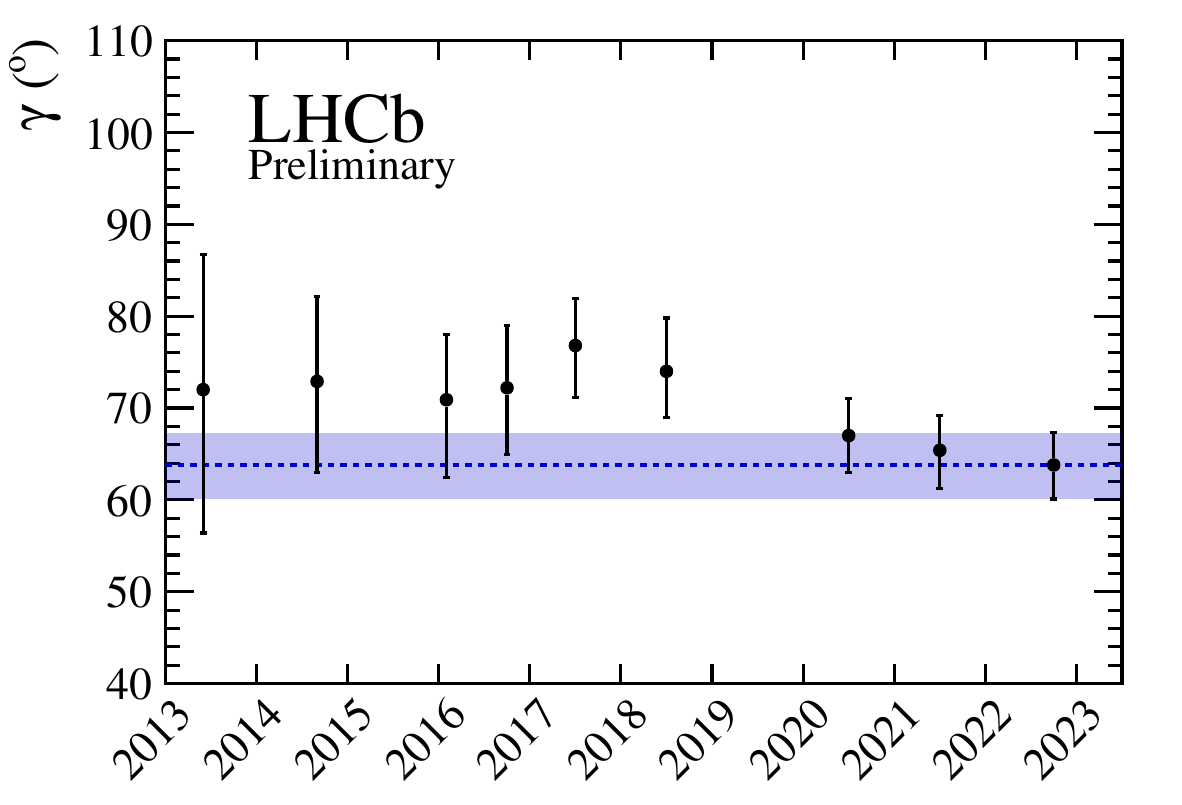}
 \end{center}
 \vspace*{-0.5cm}
 \caption{ 
    Left: the 1$-$CL value  as a function of  $\gamma$ from the
combination using inputs from $B_{s}^{0}$, $B^{0}$ and $B^{+}$ mesons and all species together; 
right: evolution of  the LHCb  $\gamma$ combination result.
   }
\label{fig_gamma_combination}
\end{figure}

A previous combination of all LHCb measurements sensitive to the angle $\gamma$ and charm mixing parameters led to the value  $\gamma = (65.4^{+3.8}_{-4.2})^{\circ}$~\cite{LHCb-PAPER-2021-033},   
which dominates the world average of direct measurements, $\gamma = (67 \pm 4)^{\circ}$~\cite{HFLAV21}.
Since then, some new measurements of $\gamma$ have emerged.
A binned analysis of  $B^{\pm} \to D[K^{\mp}\pi^{\pm}\pi^{\pm}\pi^{\mp}] h^{\pm} $ decays using  the full Run 1 and Run 2 data samples and  hadronic parameters of  $D$ decays from CLEO-c and BESIII~\cite{BESIII:2021eud} has
yielded the result $\gamma = (54.8 ^{+3.8\;+0.6\;+6.7}_{-5.8\;-0.6\;-4.3})^{\circ}$~\cite{LHCb-PAPER-2019-019}, where the first uncertainty is statistical, the second systematic and the third
due to the external inputs.
This is one of the most precise single measurements to date. A weak constraint on $\gamma$ is also obtained in  
$B^{\pm} \to D[h^{\pm} h^{\prime \mp} \pi^0] h^{\pm} $ decays~\cite{LHCb-PAPER-2021-036}. 
A new LHCb combination including these two results gives    $\gamma = (63.8^{+3.5}_{-3.7})^{\circ}$~\cite{LHCb-CONF-2022-003},  in good agreement with the indirectly determined value 
$\gamma= (65.7^{+1.3}_{-1.2})^{\circ}$~\cite{CKMfitter2015}.
Figure~\ref{fig_gamma_combination} (left) compares the constraints on $\gamma$ from decays of different $B$ species, showing a tension at $2.2\sigma$ level between \Bd and $B^+$ mesons. Figure~\ref{fig_gamma_combination} (right) shows the 
continuous improvement of the  LHCb $\gamma$ combination result in the past decade. LHCb aims to  improve the $\gamma$ precision  to  better than $1^{\circ}$ with 50 ${\rm fb}^{-1}$ from phase-I upgrade and $0.4^{\circ}$ with 300 ${\rm fb}^{-1}$  from phase-II upgrade. To achieve these goals, the  
 hadronic parameters of $D$ decays 
need be significantly improved, using more quantum-correlated $D\bar{D}$ data from BESIII and  the planned  Super $\tau$-charm factory~\cite{Cheng:2022tog}.

Belle II is starting to contribute to the $\gamma$ measurement.
Through combined analysis of  Belle and Belle II data, it has measured $\gamma = (78.4\pm 11.4 \pm 0.5  \pm 1.0  )^{\circ}$~\cite{Belle:2021efh}, and obtained \CP asymmetries in $B\to D h$ decays with $D \to K^0_S K^{\pm} \pi^{\mp}$ that can be used to constrain $\gamma$
~\cite{Belle-II:2023vtv}.
Belle II is expected to measure $\gamma$ with a precision of $1$-$2$$^{\circ}$ using  50 ${\rm ab}^{-1}$ of data~\cite{Belle-II:2010dht}.

\section{Measurement of direct \CP violation in charm decays}

\begin{figure}[tb]
\begin{center}
\includegraphics[width=0.50\linewidth]{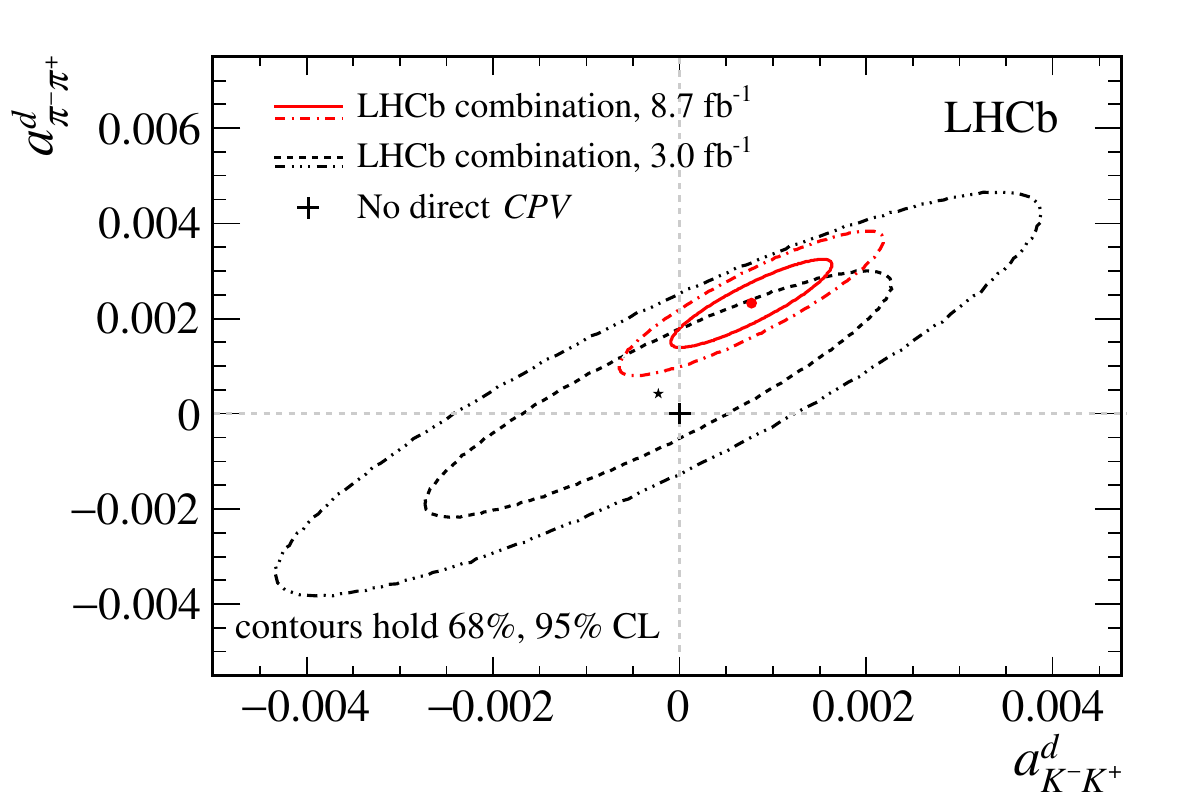}\\
 \end{center}
 \vspace*{-1.5mm}
 \caption{ Central values and confidence regions in the $a^d_{\pi^-\pi^+}$ vs $a^d_{K^-K^+}$ plane for the  LHCb results obtained with data taken during 2010--2018 and  2010--2012.}
\label{fig_D2pipi_DCPV}
\end{figure}

In 2019, LHCb observed significant \CP violation in the difference between the time-integrated \CP asymmetries in $D^0\to K^+K^-$ and $D^0\to \pi^+\pi^-$ decays, $\Delta A_{\CP}=(-15.4\pm 2.9 )\times 10^{-4}$~\cite{LHCb-PAPER-2019-006}, using the Run 2 data sample. The quantity $\Delta A_{\CP}$ could receive contributions from the decay amplitudes, the mixing and the interference of decay and mixing, but the exact breakdown is unknown. Recently, LHCb measured the time-integrated \CP asymmetry in the decay $D^0 \to K^+ K^-$ using the same data sample~\cite{LHCb-PAPER-2022-024}.
Combining this result with the   $\Delta A_{\CP}$ measurement 
determines the direct \CP asymmetries in the two decay processes.
The  obtained value  for $D^0 \to \pi^+ \pi^-$ is 
$a_{d}(\pi^+\pi^-) = (23.2\pm 6.1)\times 10^{-4}$, which is $3.8\sigma$ from zero (Figure~\ref{fig_D2pipi_DCPV}). This 
is the  first evidence  for direct \CP violation in $D^0 $ decays. 
LHCb also searched for direct \CP violation in the decays $D^0\to \pi^+\pi^-\pi^0$~\cite{LHCb-PAPER-2023-005}, $D_{(s)}^+\to K^-K^+K^+$~\cite{LHCb-PAPER-2022-042} and $D^0\to h^+h^-\mu^+\mu^-$~\cite{LHCb-PAPER-2021-035}, with no evidence found.

\section{Summary}
Using proton-proton data collected in the first two operation periods, the LHCb  experiment has  obtained high-precision measurements of \CP violation in beauty and charm decays, 
which all agree with the SM predictions. 
The Belle II experiment  is ramping up and producing interesting results.
A deeper investigation of \CP violation to fully understand its nature and origin is a long-term effort that requires synergies of LHCb, Belle II and charm experiments as well as theoretical physicists.

\bibliographystyle{LHCb}

\bibliography{main, LHCb-PAPER}

\end{document}